\documentclass[english,superscriptaddress,showpacs,showkeys]{revtex4-2}
\usepackage[T1]{fontenc}
\usepackage[utf8]{inputenc}
\usepackage{color}
\usepackage{babel}
\usepackage{array}
\usepackage{units}
\usepackage{amsmath}
\usepackage{amssymb}
\usepackage{graphicx}
\usepackage[pdfusetitle,
 bookmarks=true,bookmarksnumbered=false,bookmarksopen=true,bookmarksopenlevel=2,
 breaklinks=true,pdfborder={0 0 0},pdfborderstyle={},backref=false,colorlinks=true]
 {hyperref}
\hypersetup{
 citecolor=blue,linkcolor=blue,urlcolor=blue}

\makeatletter

\providecommand{\tabularnewline}{\\}

\makeatother

\begin{document}
\title{Coupled channels and production of near-threshold $B^{(*)}\bar{B}^{(*)}$
resonances in $e^{+}e^{-}$ annihilation}
\author{S. G. Salnikov}
\email{S.G.Salnikov@inp.nsk.su}

\affiliation{Budker Institute of Nuclear Physics, 630090, Novosibirsk, Russia}
\affiliation{Novosibirsk State University, 630090, Novosibirsk, Russia}
\author{A. E. Bondar}
\email{A.E.Bondar@inp.nsk.su}

\affiliation{Budker Institute of Nuclear Physics, 630090, Novosibirsk, Russia}
\affiliation{Novosibirsk State University, 630090, Novosibirsk, Russia}
\author{A. I. Milstein}
\email{A.I.Milstein@inp.nsk.su}

\affiliation{Budker Institute of Nuclear Physics, 630090, Novosibirsk, Russia}
\affiliation{Novosibirsk State University, 630090, Novosibirsk, Russia}
\date{\today}
\begin{abstract}
The effects of final-state interaction of hadrons, produced in $e^{+}e^{-}$
annihilation near the threshold, are discussed. If there is a loosely
bound state or a virtual state in a system of produced hadrons, then
the energy dependence of hadroproduction cross section is very strong.
Our approach is based on the use of the effective potentials accounting
for the interaction between hadrons in the final state. The cases
of a few channels with nonzero transition amplitudes between them
are considered. It is shown that these transitions drastically change
the energy dependence of the cross sections. In particular, a narrow
resonance below the threshold in one channel leads to a broad peak
in another channel. We explained the non-trivial energy dependence
of the production cross sections of $B\bar{B}$, $B^{*}\bar{B}$,
$B\bar{B}^{*}$, $B^{*}\bar{B}^{*}$ near the thresholds in $e^{+}e^{-}$
annihilation and obtained good agreement between our predictions and
the experimental data available.
\end{abstract}
\maketitle

\section{Introduction}

\global\long\def\im#1{\qopname\relax{no}{Im}#1}%

Typical values of potentials, describing interaction of hadrons, are
relatively large (hundreds of MeV), and the radii of these potentials
are about 1~fm. Therefore, a loosely bound state may exist with the
modulus of binding energy much less than a depth of the potential
well. In this case, the scattering length $\xi$ is positive and considerably
exceeds the radius $a$ of the potential. The binding energy $\varepsilon$
is expressed via the scattering length as $\varepsilon=-1/M\xi^{2}$,
where $M$ is the mass of hadrons. It is also possible that there
is no bound state, but a slight increase in the potential depth leads
to appearance of such a state (so-called virtual state). In this case
the scattering length is negative, and its absolute value is much
larger than~$a$. The energy of the virtual state is defined as $\varepsilon=1/M\xi^{2}$.
In both cases (bound or virtual state) a resonance is observed in
the hadron scattering cross section.

The process of heavy meson or baryon production in $e^{+}e^{-}$ annihilation
starts with heavy quark-antiquark pair production at small distances
of the order of $1/\sqrt{s}$. At larger distances the final state
consists of mesons or baryons described by the wave function $\psi(\boldsymbol{r})$.
Therefore, the production cross section of heavy hadrons is proportional
to $\left|\psi(0)\right|^{2}$ for a pair in the s-wave, $\left|\psi'(0)\right|^{2}$
for a pair in the p-wave, etc. (see Refs.~\citep{Dmitriev2006,dmitriev2007final}).
For a low-energy bound or virtual state, the absolute value of the
wave function or its derivative inside the potential well is much
larger than the corresponding valued in the absence of interaction
potential. As a result, the hadroproduction cross sections increase
near the thresholds of the corresponding processes. Note that the
shape of near-threshold resonances differs drastically from that obtained
using the commonly used Breit-Wigner formula.

At present, near-threshold resonances in $e^{+}e^{-}$ annihilation
have been observed in many processes: $e^{+}e^{-}\to p\bar{p}$~\citep{Aubert2006,Lees2013,Akhmetshin2016,Ablikim2020,Ablikim2021b,Akhmetshin2019,Ablikim2015,Ablikim2019},
$e^{+}e^{-}\to n\bar{n}$~\citep{Achasov2014,Ablikim2021f,Achasov2022},
$e^{+}e^{-}\to\Lambda\bar{\Lambda}$~\citep{Aubert2007,Ablikim2018,Ablikim2019c,Ablikim2023},
$e^{+}e^{-}\to\Lambda_{c}\bar{\Lambda}_{c}$~\citep{Pakhlova2008,Ablikim2018b},
$e^{+}e^{-}\to B\bar{B}$~\citep{Aubert2009}, and others. In all
these cases, the shapes of near-threshold resonances differ from that
described by the Breit-Wigner formula and differ also from each other.
The shapes of resonances are determined by many factors: isotopic
structure of the produced states, angular momentum, magnitude of tensor
forces, the Coulomb interaction of charged particles, and others.
For instance, tensor forces are responsible for transitions between
the states with different angular momenta and affect the ratio of
hadron electromagnetic form factors. The account for the final-state
interaction of produced hadrons allows one to describe all experimental
data within the same approach (see Refs.~\citep{Haidenbauer2014,Haidenbauer2016,Haidenbauer2021,Milstein2021,Milstein2022a,Milstein2022,Milstein2022c,Salnikov2023}
and references therein).

In addition to a non-trivial behavior of the hadroproduction cross
sections near the corresponding thresholds, the production cross sections
of light mesons reveal a strong energy dependence near the same thresholds.
This behavior is due to production of virtual hadrons below and above
threshold with subsequent annihilation into light mesons. Examples
of such processes are $e^{+}e^{-}\to6\pi$~\citep{Aubert2006a,Akhmetshin2013,Lukin2015,Akhmetshin2019},
$e^{+}e^{-}\to K^{+}K^{-}\pi^{+}\pi^{-}$~\citep{Aubert2005,Aubert2007c,Akhmetshin2019},
$J/\psi\to\gamma\eta'\pi^{+}\pi^{-}$~\citep{Ablikim2016}, and others.
The unusual energy dependence of the cross sections of these processes
have been explained by the interaction of virtual hadrons in the intermediate
state~\citep{Haidenbauer2015,Dmitriev2016,Milstein2022c}.

In this work, we use the final-state interaction approach to describe
the production of $B\bar{B}$, $B\bar{B}^{*}$, $B^{*}\bar{B}$, $B^{*}\bar{B}^{*}$
in $e^{+}e^{-}$ annihilation near the thresholds of these processes.
The corresponding experimental data are published in Refs.~\citep{Aubert2009,Dong2020,Mizuk2021,Bertemes2023}.
In Refs.~\citep{Milstein2021,Bondar2022}, the final-state interaction
approach was used to study the effects of isotopic invariance violation
in the production of $B\bar{B}$ and $B^{*}\bar{B}^{*}$ near the
thresholds. Since the states $B\bar{B}$, $B\bar{B}^{*}$, $B^{*}\bar{B}^{*}$,
produced in $e^{+}e^{-}$ annihilation, have identical quantum numbers
$J^{PC}=1^{--}$, then the amplitudes of transitions between these
states are not zero. These transitions are irrelevant to the effects
of isotopic invariance violation, since only narrow region close to
the peak of resonance is essential. However, the account for the transition
amplitudes is crucial for the description of the entire energy region
near the thresholds of coupled channels. This is the main goal of
our paper.

The paper is organized as follows. In Section~\ref{sec:Simple-model},
using a simple potential model of rectangular wells, we elucidate
the main features of the resonance production cross sections near
the thresholds in the cases of one channel, two channels, and three
channels. In Section~\ref{sec:Production-of-}, the our approach
is applied to description of the experimental data on the cross sections
of $B^{(*)}\bar{B}^{(*)}$ pair production near the thresholds. The
main results obtained are summarized in Conclusion.

\section{Simple model}\label{sec:Simple-model}

In this section, using a simple approach, we discuss the cross sections
of $e^{+}e^{-}$ annihilation into hadron-antihadron pairs near the
thresholds of the processes. This approach allows one to understand
the origin of the enhancement of the cross sections and to achieve
good agreement with available experimental data as well. We consider
the cases of one, two, and three channels in the final state having
the same quantum numbers $J^{PC}$ and close thresholds. We show that
the shape of the cross section drastically depends on the number of
channels due to transitions between them.

\subsection{One channel}

Let us consider only one channel when the radial wave function $\psi(r)$
of produced hadrons satisfies the Schrödinger equation
\begin{equation}
\left(p_{r}^{2}+MV+\frac{l(l+1)}{r^{2}}-k^{2}\right)\psi(r)=0\,,\label{eq:Schro}
\end{equation}
where $(-p_{r}^{2})$~is the radial part of the Laplacian operator,
$M$~is the hadron mass, $k=\sqrt{ME}$, $E$~is the kinetic energy
of a pair counted from the reaction threshold, and $V$~is the interaction
potential. We are interested in regular solution $\psi_{R}(r)$ with
asymptotic behavior 
\begin{equation}
\psi_{R}(r\to\infty)=\frac{1}{2ikr}\left(S\chi^{+}-\chi^{-}\right),\qquad\chi^{\pm}=\exp\left[\pm i\left(kr-\pi l/2\right)\right].
\end{equation}
Cross section of pair production in the state with angular momentum
$l$ is given by the relation (see, e.g., Refs.~\citep{Milstein2021,Milstein2022c}
and references therein)
\begin{equation}
\sigma=\frac{2\pi\beta\alpha^{2}}{s}\left|g\,\psi_{R}^{(l)}(0)\right|^{2}\,,\label{eq:sig1}
\end{equation}
where $\beta=k/M$~is the hadron velocity, $s=\left(2M+E\right)^{2}$,
and $\psi_{R}^{(l)}(r)=\left(\partial/\partial r\right)^{l}\psi_{R}(r)$.
The coefficient $g$ is related to the amplitude of pair production
at small distances $\sim1/\sqrt{s}$ and can be considered as an energy
independent constant.

The strong energy dependence of the cross section near the threshold
appears if there is a bound state with a small value of binding energy
or a low-energy virtual state. In the latter case a small increase
of the potential depth leads to appearance of a bound state with a
small binding energy. To illustrate this point, let us consider a
rectangular potential well $V(r)=U\,\theta(a-r)$, where $U<0$, and
$\theta(x)$~is the Heaviside function. For $l=0$, Eq.~(\ref{eq:Schro})
can easily be solved analytically. As a result, the wave function
at the origin reads
\begin{equation}
\psi_{R}(0)=\frac{q\,e^{-ika}}{q\cos\left(qa\right)-ik\sin\left(qa\right)}\,,\qquad q=\sqrt{M(E-U)}\,.\label{eq:psi0}
\end{equation}
Near the threshold, we have $k\ll q$. Then, the cross section~(\ref{eq:sig1})
is enhanced if
\begin{equation}
q_{0}a\approx\pi\left(n+\frac{1}{2}\right)+\delta\,,\qquad\left|\delta\right|\ll1\,,\label{eq:delta}
\end{equation}
where $q_{0}=\sqrt{M\left|U\right|}$, and $n$~is an integer. Under
this condition a low-energy bound or virtual state exists in the potential
well. For $\left|\delta\right|\ll1$, the scattering length is $\xi=1/q_{0}\delta$,
where $\delta>0$ for the bound state with the binding energy $\varepsilon=-1/M\xi^{2}$,
and $\delta<0$ for the virtual state with the energy $\varepsilon=1/M\xi^{2}$.
In both cases $\left|\varepsilon\right|\ll\left|U\right|$.

By means of Eqs.~(\ref{eq:psi0}) and~(\ref{eq:delta}) the expression
for $\left|\psi_{R}(0)\right|^{2}$ can be simplified (see Ref.~\citep{Salnikov2023}):
\begin{align}
 & \left|\psi_{R}(0)\right|^{2}=\frac{q^{2}}{q^{2}\cos^{2}\left(qa\right)+k^{2}\sin^{2}\left(qa\right)}\approx\frac{\gamma\left|U\right|}{\left(E+\varepsilon_{0}\right)^{2}+\gamma E}\,,\nonumber \\
 & \gamma=4\kappa^{2}\left|U\right|\,,\qquad\varepsilon_{0}=2\kappa\,\delta\left|U\right|\,,\qquad\kappa=\frac{1}{\pi(n+1/2)}\,.
\end{align}
Corresponding energy dependence of the cross section~(\ref{eq:sig1})
is equivalent to the Flatté formula~\citep{Flatte1976}, which is
expressed in terms of the scattering length and the effective radius
of interaction. Note that for the rectangular potential well the latter
equals to the radius~$a$ of the potential. One can easily verify
that the precise and approximate formulas for the cross section are
in good agreement with each other for $\left|\delta\right|\ll1$ and
$E\lesssim\varepsilon_{0}\ll\left|U\right|$. Note that $\left|\varepsilon_{0}\right|\gg\left|\varepsilon\right|$
for both bound and virtual states, namely $\varepsilon_{0}\approx2\left|\varepsilon\right|\xi/a$.
However, the position of peak in the cross section, which is proportional
to $\sqrt{E}\left|\psi_{R}(0)\right|^{2}$, is located at energy $E\approx\left|\varepsilon\right|$
for both bound and virtual states.

Thus, the description of near-threshold resonances by means of Flatté
formula or by model potential are equivalent. However, the approach
based on the use of model potential significantly simplifies consideration
in the case of multiple coupled channels. The effective potential
method is very convenient for account of the Coulomb interaction between
hadrons. In is especially powerful for calculation of the total cross
section of a process which includes an annihilation of real or virtual
hadron pairs into light mesons, see, e.g., Ref.~\citep{Milstein2022c}
and references therein.

\begin{figure}
\centering
\includegraphics[totalheight=5.7cm]{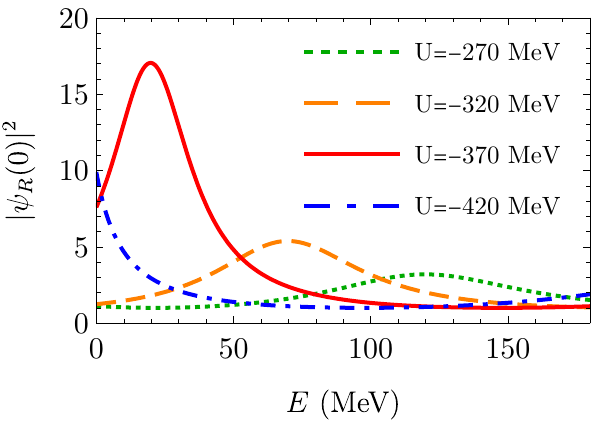}\hspace*{\fill}\includegraphics[totalheight=5.7cm]{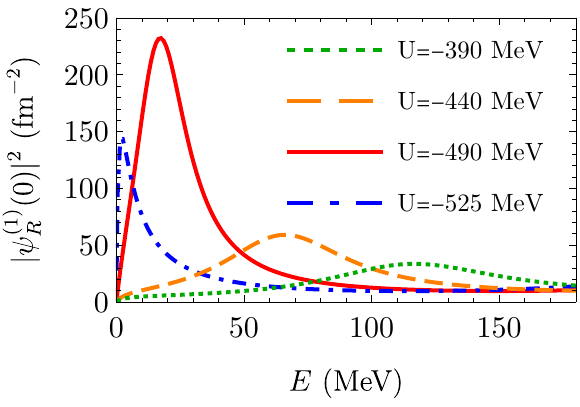}

\caption{The energy dependence of $\left|\psi_{R}(0)\right|^{2}$ for $l=0$
(left) and $\left|\psi_{R}^{(1)}(0)\right|^{2}$ for $l=1$ (right).
The radius $a=\unit[1.5]{fm}$ and a few values of $U$ are used.}\label{fig:psi1}
\end{figure}

As an example, we set $M=M_{B}$, where the mass of neutral $B$ meson
is $M_{B}=\unit[5279.65]{MeV}$. The energy dependence of $\left|\psi_{R}(0)\right|^{2}$
with the wave function given in Eq.~(\ref{eq:psi0}) is shown in
Fig.~\ref{fig:psi1} for $a=\unit[1.5]{fm}$ and a few values of
potential well depths. For such value of $a$, the bound state with
$n=3$ appears at $U=\unit[-396]{MeV}$. At $U=\unit[-420]{MeV}$
there is a bound state with binding energy $\varepsilon=\unit[-12]{MeV}$.
For $U=\unit[-370]{MeV}$, there is a virtual state, which results
in the peak in the energy dependence of $\left|\psi_{R}(0)\right|^{2}$
at $E\approx\unit[20]{MeV}$. For smaller values of the potential
well depth, the peak becomes less pronounced.

For $l=1$, the cross section of hadron production~(\ref{eq:sig1})
is proportional to the factor $\left|\psi_{R}^{(1)}(0)\right|^{2}$.
The analytical form of this quantity is given in Eq.~(8) of Ref.~\citep{Milstein2021}.
The energy dependence of $\left|\psi_{R}^{(1)}(0)\right|^{2}$ for
$a=\unit[1.5]{fm}$ and a few potential well depths is shown in Fig.~\ref{fig:psi1}.
For $l=1$ and $U=\unit[-504]{MeV}$, a loosely bound state, corresponding
to $n=4$, appears. At $U=\unit[-525]{MeV}$ there is a low-energy
bound state with $\varepsilon=\unit[-4]{MeV}$. For $U=\unit[-490]{MeV}$,
there is a virtual state, which leads to a peak in $\left|\psi_{R}(0)\right|^{2}$
at $E\approx\unit[20]{MeV}$. As for $l=0,$ for smaller values of
$\left|U\right|$, the peak becomes less lower and wider. Qualitatively,
the pictures for $l=0$ and $l=1$ look very similar.

\begin{figure}
\centering
\includegraphics[totalheight=6cm]{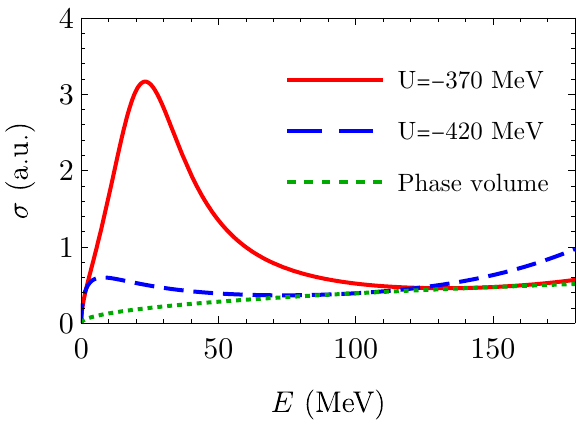}\hspace*{\fill}\includegraphics[totalheight=6cm]{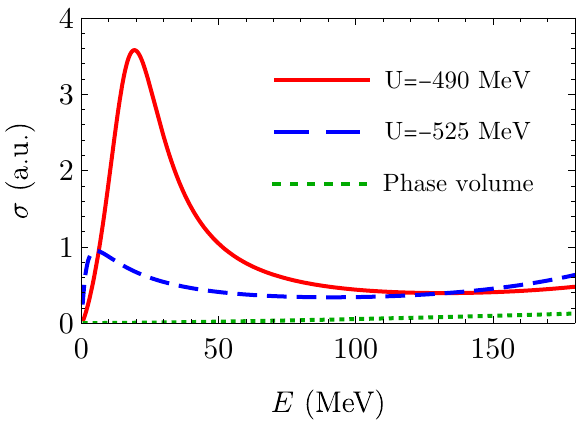}

\caption{The cross section of hadron pair production for $l=0$ (left) and
$l=1$ (right) at $a=\unit[1.5]{fm}$ and various depths of the potential
well. The cross section without final-state interaction corresponds
to the energy dependence of the phase volume.}\label{fig:sig1}
\end{figure}

The energy dependence of the cross section within our simple model
is shown in Fig.~\ref{fig:sig1} for a few values of the potential
parameters. We consider the cases when a low-energy bound or virtual
state with $l=0$ or $l=1$ exists. It is seen that for some depth
of the potential well there is a near-threshold peak in the cross
section. As mentioned above, the position of this peak is determined
by the energy of the virtual state, while the shape of this peak depends
also on the effective radius of interaction. The near-threshold peaks
turned out to be quite smooth and the next peaks in the cross section
are located much higher in energy.

Let us include into consideration the possibility for produced hadron
pair to annihilate into other particles. The number of different final
states in this inelastic process can be huge and the probability of
such annihilation can be significant. Inelastic processes are usually
taken into account by means of so-called optical potentials in the
same way as it is done in nuclear physics. These optical potentials
contain not only the real part, but also the imaginary part. We refer
to the cross section of real pair production as the elastic cross
section, $\sigma_{\textrm{el}}$, which is given by Eq.~(\ref{eq:sig1}).
The inelastic cross section, $\sigma_{\textrm{in}}$, corresponds
to the process, in which a produced virtual hadron pair annihilates
into the final state. Note that the inelastic cross section, as well
as the total cross section $\sigma_{\textrm{tot}}=\sigma_{\textrm{el}}+\sigma_{\textrm{in}}$,
reveals a strong energy dependence near the threshold of real hadron
pair production. This effect was discovered experimentally (see, e.g.,
Refs.~\citep{Aubert2006a,Akhmetshin2013,Lukin2015,Akhmetshin2019,Aubert2005,Aubert2007c})
and then have been explained theoretically (see Ref.~\citep{Milstein2022c}
and references therein).

The total cross section is expressed via the Green's function $\mathcal{D}(r,r'|E)$
of the Schrödinger equation (see Ref.~\citep{Milstein2022c} and
references therein):
\begin{equation}
\sigma_{\textrm{tot}}=\frac{2\pi\alpha^{2}}{Ms}\left|g\right|^{2}\im{\mathcal{D}(0,0|E)}\,,
\end{equation}
where $\mathcal{D}(r,r'|E)$ satisfies the equation
\begin{equation}
\left(p_{r}^{2}+MV+\frac{l(l+1)}{r^{2}}-k^{2}\right)\mathcal{D}(r,r'|E)=\frac{1}{rr'}\delta(r-r')\,.
\end{equation}
The Green's function can be expressed in terms of the regular solution
$\psi_{R}(r)$ and the non-regular solution $\psi_{N}(r)$ of Eq.~(\ref{eq:Schro}),
having the asymptotics
\begin{equation}
\psi_{N}(r\to\infty)=\frac{1}{kr}\,\chi^{+}\,.
\end{equation}
We have
\begin{equation}
\mathcal{D}(r,r'|E)=k\left(\psi_{R}(r)\psi_{N}(r')\theta(r'-r)+\psi_{R}(r')\psi_{N}(r)\theta(r-r')\right).
\end{equation}
For $l=0$ one can easily obtain the analytical expression for the
imaginary part of the Green's function
\begin{equation}
\im{\mathcal{D}(0,0|E)}=\im{\left[q\,\frac{q\sin\left(qa\right)+ik\cos\left(qa\right)}{q\cos\left(qa\right)-ik\sin\left(qa\right)}\right]}.
\end{equation}
Note that for complex potential $U$ the quantity $q$ is also complex.
For $\im U=0$, the cross section $\sigma_{\textrm{tot}}$ above the
threshold coincides with the elastic cross section~$\sigma_{\textrm{el}}$.
Below the threshold $\sigma_{\textrm{tot}}$ contains contributions
of possible bound states.

\begin{figure}
\centering
\includegraphics[totalheight=6cm]{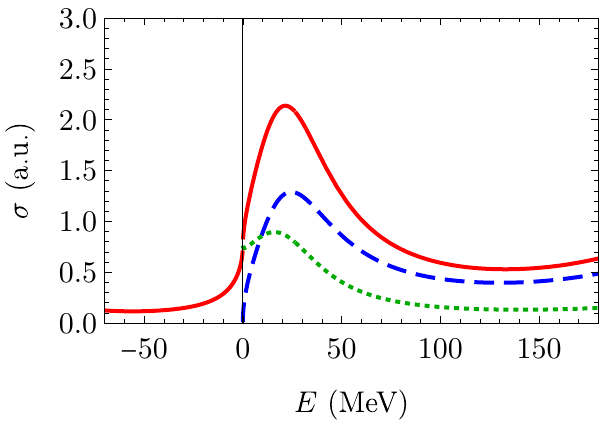}\hspace*{\fill}\includegraphics[totalheight=6cm]{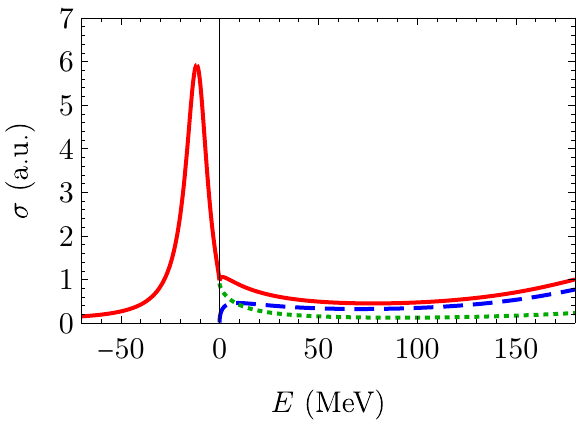}

\caption{The cross sections of hadron pair production for $l=0$ and $a=\unit[1.5]{fm}$.
The depths of the optical potential are $U=\unit[\left(-370-10i\right)]{MeV}$
(left) and $U=\unit[\left(-420-10i\right)]{MeV}$ (right). The dashed
line corresponds to $\sigma_{\textrm{el}}$, the dotted line to $\sigma_{\textrm{in}}$,
and the solid line to $\sigma_{\textrm{tot}}$.}\label{fig:sig1tot}
\end{figure}

The energy dependence of the cross sections $\sigma_{\textrm{el}}$,
$\sigma_{\textrm{in}}$ and $\sigma_{\textrm{tot}}$ is shown in Fig.~\ref{fig:sig1tot}
for $\im{U=\unit[-10]{MeV}}$. The left and right plots correspond,
respectively, to the cases when a virtual state and a bound state
exist. Though the imaginary part of potential is small, the elastic
cross section is significantly smaller than that for $\im U=0$ (cf.
Fig.~\ref{fig:sig1}). At the same time, above the threshold the
inelastic cross section $\sigma_{\textrm{in}}$ becomes comparable
to $\sigma_{\textrm{el}}$. Below the threshold, a peak is observed
if a bound state exists. The width and height of this peak are determined
by the imaginary part of the optical potential. With vanishing of
the imaginary part of the potential, this peak turns into a $\delta$-function
at the energy of a bound state. Naturally, in the case of a virtual
level, a peak below the threshold does not occur.

\subsection{Two channels}

Let us consider now a two-channel process in which hadron pairs appear
with the same quantum numbers $J^{PC}=1^{--}$, close masses but different
spins. To be specific, we will talk about pairs in the states $X=B\bar{B}$
and $Y=\left(B^{*}\bar{B}-B\bar{B}^{*}\right)/\sqrt{2}$. Since the
spins of $B$ and $B^{*}$ mesons are different, the interaction potential
$V_{XX}$ between $B$ and $\bar{B}$ differs slightly from the interaction
potential $V_{YY}$ between $B^{*}$ and~$\bar{B}$. In addition,
identical quantum numbers $J^{PC}$ make transition between the $X$
and $Y$ states to be possible. The corresponding off-diagonal matrix
element $V_{XY}$ is small compared to the potentials $V_{XX}$ and
$V_{YY}$. However, the value of $V_{XY}$ should be compared not
with these potentials, but with the energies of the real or virtual
states. We show that in this case the energy dependence of the pair
production cross sections differs noticeably from the case of $V_{XY}=0$.

In the near-threshold energy region the wave function of produced
hadron pairs satisfies the coupled-channel Schrödinger equation
\begin{equation}
\left(p_{r}^{2}+M_{B}\mathcal{V}+\frac{l(l+1)}{r^{2}}-\mathcal{K}^{2}\right)\Psi(r)=0\,,\qquad\mathcal{K}^{2}=\begin{pmatrix}k_{X}^{2} & 0\\
0 & k_{Y}^{2}
\end{pmatrix},\qquad\mathcal{V}=\begin{pmatrix}V_{XX} & V_{XY}\\
V_{XY} & V_{YY}
\end{pmatrix}.\label{eq:Schro2}
\end{equation}
Here $k_{X}=\sqrt{M_{B}E}$, $k_{Y}=\sqrt{M_{B}\left(E-\Delta\right)}$,
$\Delta=M_{B^{*}}-M_{B}=\unit[45.97]{MeV}$, $M_{B}=\unit[5279.65]{MeV}$
and $M_{B^{*}}=\unit[5325.62]{MeV}$~are the masses of corresponding
neutral mesons. In Eq.~(\ref{eq:Schro2}), the wave function $\Psi=\left(u,v\right)^{T}$
consists of a radial wave function $u$ of the $X$ state and the
radial wave function $v$ of the $Y$ state. The superscript $T$
denotes the transposition. In following, we assume that the angular
momentum of pairs is $l=1$, since it corresponds to the process of
$B$-meson pair production in $e^{+}e^{-}$ annihilation.

The cross sections of meson production in the case of two channels
are
\begin{align}
 & \sigma_{X}=\frac{2\pi\beta_{X}\alpha^{2}}{s}\left|g_{X}u_{1R}^{(1)}(0)+g_{Y}v_{1R}^{(1)}(0)\right|^{2},\nonumber \\
 & \sigma_{Y}=\frac{2\pi\beta_{Y}\alpha^{2}}{s}\left|g_{X}u_{2R}^{(1)}(0)+g_{Y}v_{2R}^{(1)}(0)\right|^{2},\label{eq:sig2}
\end{align}
where $\beta_{X}=k_{X}/M_{B}$, and $\beta_{Y}=k_{Y}/M_{B}$. Here
$u_{1R}$, $v_{1R}$, $u_{2R}$, and $v_{2R}$~are the corresponding
components of the regular solutions $\Psi_{1R}$ and $\Psi_{2R}$
of Eq.~(\ref{eq:Schro2}). These solutions have the asymptotic forms
at $r\to\infty$
\begin{align}
 & \Psi_{1R}=\frac{1}{2ik_{X}r}\left(S_{11}\chi_{X}^{+}-\chi_{X}^{-},\,S_{12}\chi_{Y}^{+}\right)^{T},\nonumber \\
 & \Psi_{2R}=\frac{1}{2ik_{Y}r}\left(S_{21}\chi_{X}^{+},\,S_{22}\chi_{Y}^{+}-\chi_{Y}^{-}\right)^{T},\nonumber \\
 & \chi_{X}^{\pm}=\exp\left[\pm i\left(k_{X}r-\pi/2\right)\right],\qquad\chi_{Y}^{\pm}=\exp\left[\pm i\left(k_{Y}r-\pi/2\right)\right].
\end{align}
Complex coefficients $g_{X}$ and $g_{Y}$~are related to the amplitudes
of corresponding meson pair production at small distances, and $S_{ij}$~are
some constants.

\begin{figure}
\centering
\includegraphics[totalheight=6cm]{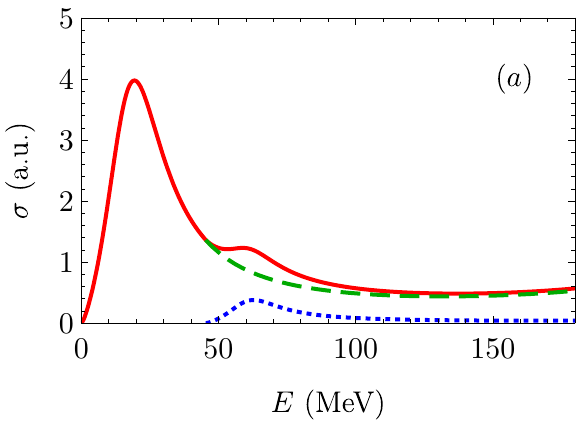}

\includegraphics[totalheight=6cm]{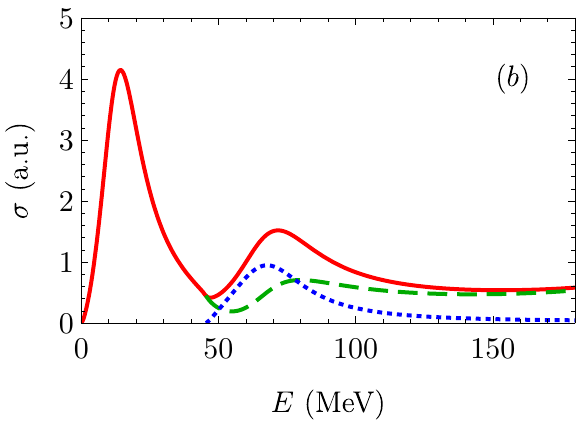}\hspace*{\fill}\includegraphics[totalheight=6cm]{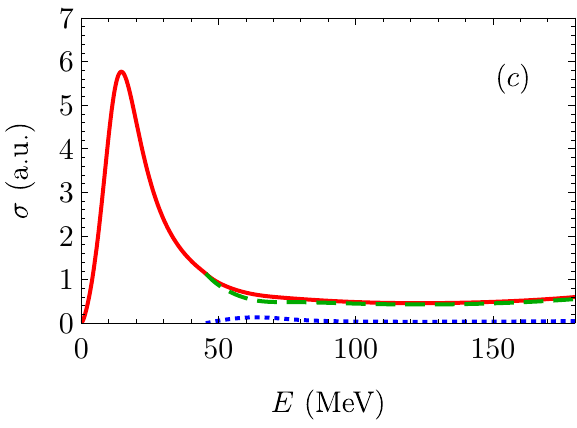}

\caption{The energy dependence of the cross sections $\sigma_{X}$ (dashed
line), $\sigma_{Y}$ (dotted line) and their sum (solid line). The
parameters of the model are $U_{XX}=U_{YY}=-\unit[490]{MeV}$, $a=\unit[1.5]{fm}$,
$g_{X}=\unit[1]{fm}$, $g_{Y}=\unit[0.3]{fm}$. On plot (a) off-diagonal
potential $U_{XY}=0$, on plot (b) $U_{XY}=\unit[20]{MeV}$, and on
plot (c) $U_{XY}=-\unit[20]{MeV}$.}\label{fig:sig2-1}
\end{figure}

In order to demonstrate how the interaction between channels affect
the energy dependence of the cross sections, let us consider a few
examples. We choose the parametrization of the potential in the form
of rectangular wells with the same radius but different depths:
\begin{equation}
\mathcal{V}=\begin{pmatrix}U_{XX} & U_{XY}\\
U_{XY} & U_{YY}
\end{pmatrix}\theta(a-r)\,,
\end{equation}
where the matrix elements $U_{ij}$ and the radius $a$~are some
constants. Below we present the numerical results of our investigations
within such a model. Let us first consider the values of $U_{ij}$
and $a$ that lead to low-energy virtual states in both channels.
The corresponding cross sections $\sigma_{X}$, $\sigma_{Y}$, and
their sum are shown in Fig.~\ref{fig:sig2-1} as a functions of energy.
The plot~\ref{fig:sig2-1}a corresponds to $U_{XY}=0$, the plot~\ref{fig:sig2-1}b
to $U_{XY}=\unit[20]{MeV}$, and the plot~\ref{fig:sig2-1}c to $U_{XY}=-\unit[20]{MeV}$.
In all three cases we use $g_{X}=\unit[1]{fm}$ and $g_{Y}=\unit[0.3]{fm}$.
For zero $U_{XY}$, there is no pronounced dip in the total cross
section, as well as for $U_{XY}=\unit[-20]{MeV}$. However, at $U_{XY}=\unit[20]{MeV}$
the peaks in the cross section become more noticeable, and the cross
section between peaks drops almost to zero. In the latter case there
is also a dip in the cross section $\sigma_{X}$ near the threshold
of production of~$Y$. Such a difference in the behavior of the cross
sections is due to the interference of different terms in Eq.~(\ref{eq:sig2}).
Note that the cross sections are invariant under the simultaneous
change of the signs of $U_{XY}$ and $g_{Y}$.

\begin{figure}
\centering
\includegraphics[totalheight=6cm]{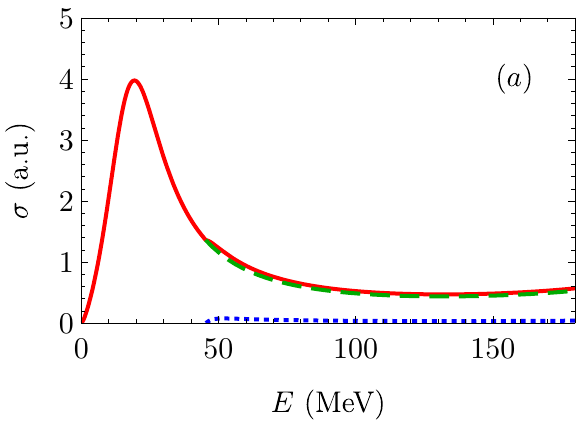}

\includegraphics[totalheight=6cm]{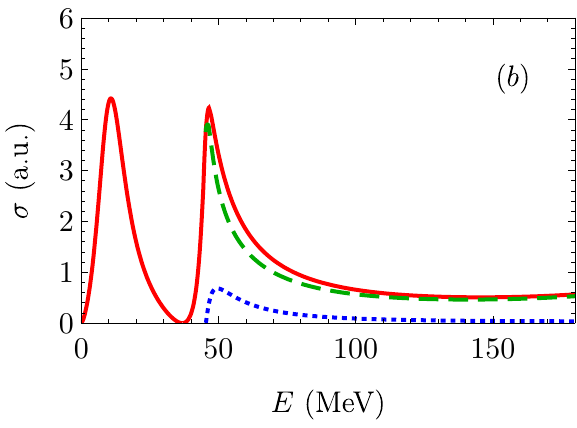}\hspace*{\fill}\includegraphics[totalheight=6cm]{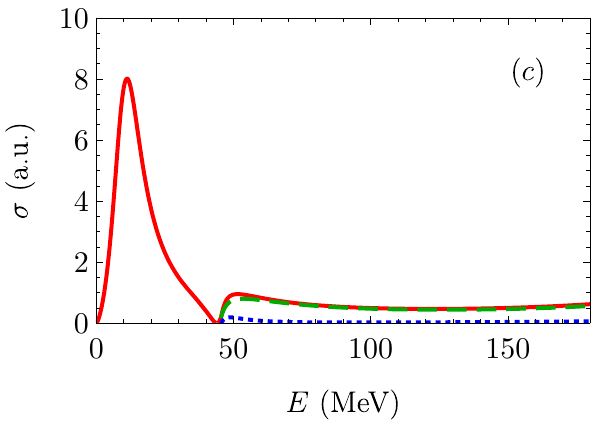}

\caption{The energy dependence of the cross sections $\sigma_{X}$ (dashed
line), $\sigma_{Y}$ (dotted line) and their sum (solid line). The
parameters of the model are $U_{XX}=-\unit[490]{MeV}$, $U_{YY}=-\unit[525]{MeV}$,
$a=\unit[1.5]{fm}$, $g_{X}=\unit[1]{fm}$, $g_{Y}=\unit[0.3]{fm}$.
On plot (a) off-diagonal potential $U_{XY}=0$, on plot (b) $U_{XY}=\unit[20]{MeV}$,
and on plot (c) $U_{XY}=-\unit[20]{MeV}$.}\label{fig:sig2-2}
\end{figure}

Now we pass to the existence of a virtual state in the $X$ channel
and a low-energy bound state in the $Y$ channel (see Fig.~\ref{fig:sig2-2}).
For $U_{XY}=0$, the bound state does not manifest itself, and the
cross section $\sigma_{Y}$ is small. However, when a small value
of $V_{XY}$ is introduced, the behavior of the cross section changes
significantly. This is due to the transition of the bound state in
the $Y$ channel into the $X$ channel with subsequent decay into
$B$ mesons. As a result, a resonant structure appears in the $B\bar{B}$
production cross section.

\begin{figure}
\centering
\includegraphics[totalheight=6cm]{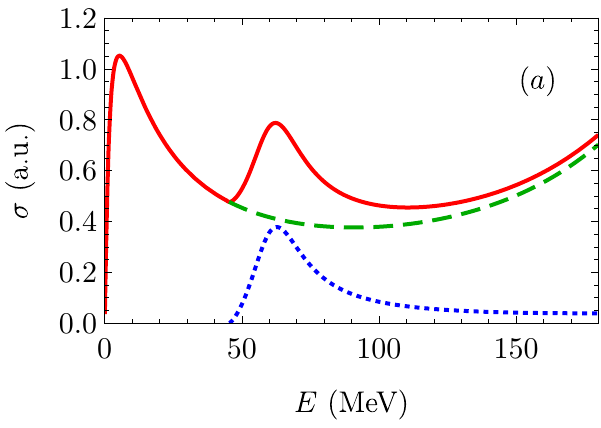}

\includegraphics[totalheight=6cm]{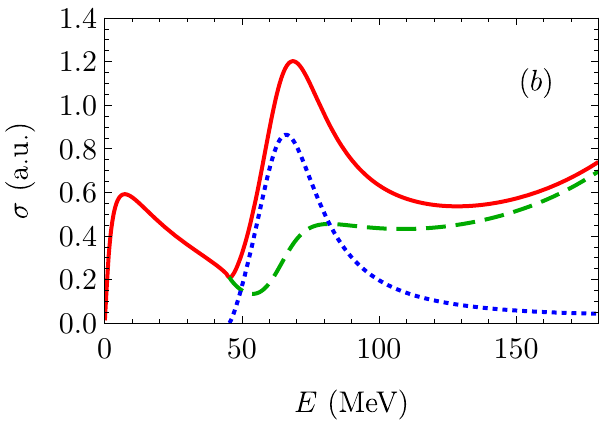}\hspace*{\fill}\includegraphics[totalheight=6cm]{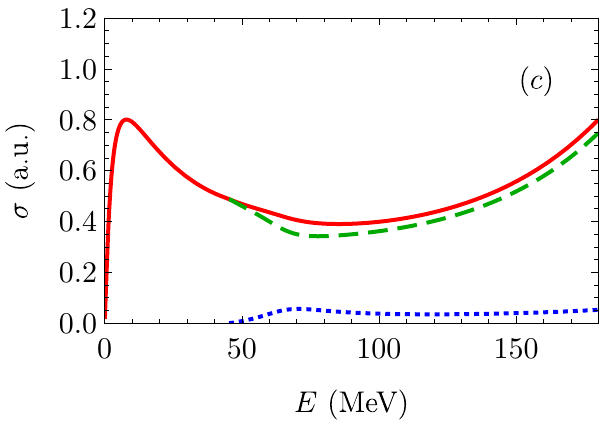}

\caption{The energy dependence of the cross sections $\sigma_{X}$ (dashed
line), $\sigma_{Y}$ (dotted line) and their sum (solid line). The
parameters of the model are $U_{XX}=-\unit[525]{MeV}$, $U_{YY}=-\unit[490]{MeV}$,
$a=\unit[1.5]{fm}$, $g_{X}=\unit[1]{fm}$, $g_{Y}=\unit[0.3]{fm}$.
On plot (a) off-diagonal potential $U_{XY}=0$, on plot (b) $U_{XY}=\unit[20]{MeV}$,
and on plot (c) $U_{XY}=-\unit[20]{MeV}$.}\label{fig:sig2-3}
\end{figure}

\begin{figure}
\centering
\includegraphics[totalheight=6cm]{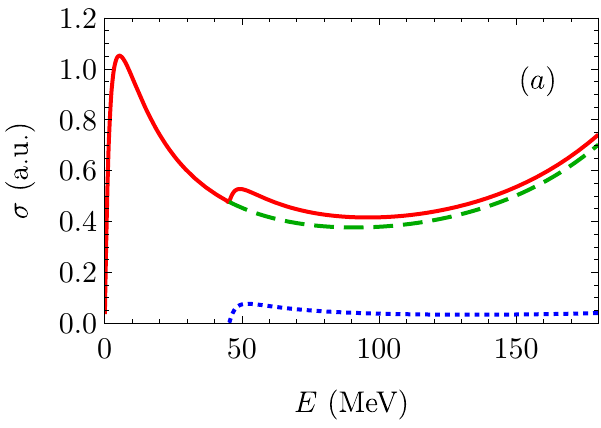}

\includegraphics[totalheight=6cm]{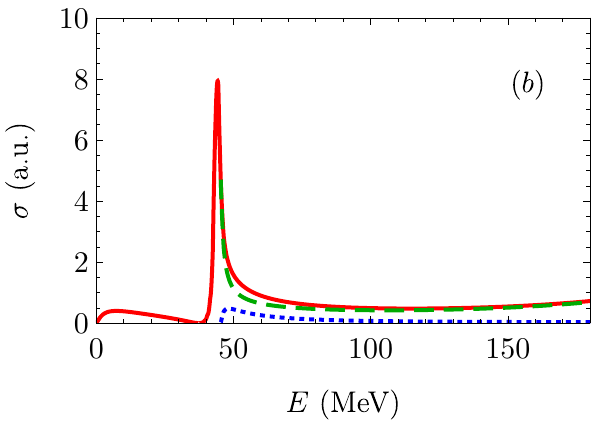}\hspace*{\fill}\includegraphics[totalheight=6cm]{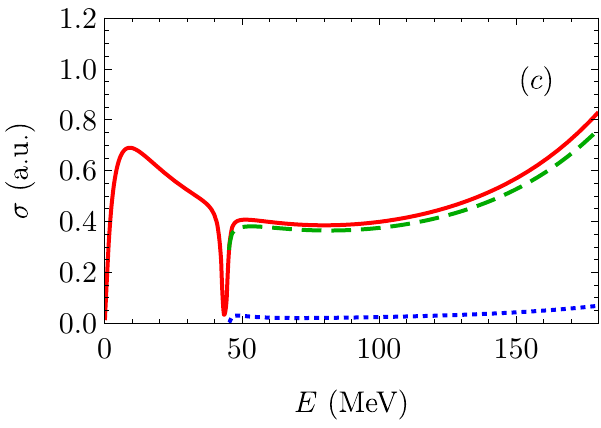}

\caption{The energy dependence of the cross sections $\sigma_{X}$ (dashed
line), $\sigma_{Y}$ (dotted line) and their sum (solid line). The
parameters of the model are $U_{XX}=U_{YY}=-\unit[525]{MeV}$, $a=\unit[1.5]{fm}$,
$g_{X}=\unit[1]{fm}$, $g_{Y}=\unit[0.3]{fm}$. On plot (a) off-diagonal
potential $U_{XY}=0$, on plot (b) $U_{XY}=\unit[20]{MeV}$, and on
plot (c) $U_{XY}=-\unit[20]{MeV}$.}\label{fig:sig2-4}
\end{figure}

For a low-energy bound state in the $X$ channel and a virtual or
a bound state in the $Y$ channel, the energy dependence of the cross
sections are shown in Figs.~\ref{fig:sig2-3} and~\ref{fig:sig2-4}.
In both cases there is no pronounced peak above the threshold of $X$
state production. However, a peak above the threshold of $Y$ state
production can be seen in Fig.~\ref{fig:sig2-3}. This peak, corresponding
to the virtual state, becomes more noticeable for one sign of $U_{XY}$
and almost disappears for another sign. If there is a bound state
in the $Y$ channel, then there are no significant peaks in the cross
section at $U_{XY}=0$. However, for a small $U_{XY}$, a resonant
structure manifests itself at the energy of a bound state. For one
sign of the potential $U_{XY}$, the is a narrow and high peak in
the cross section, and for another sign there is a sharp dip.

Similar to one channels problem, we introduce the imaginary part of
the optical potential. There are two elastic cross sections, $\sigma_{\textrm{el},X}$
and $\sigma_{\textrm{el},Y}$, which are given by Eq.~(\ref{eq:sig2}).
There is also an inelastic cross section $\sigma_{\textrm{in}}$ corresponding
to the processes of annihilation of virtual $X$ and $Y$ states into
other particles. The total cross section is a sum $\sigma_{\textrm{tot}}=\sigma_{\textrm{el},X}+\sigma_{\textrm{el},Y}+\sigma_{\textrm{in}}$
and are expressed via the Green's function $\mathcal{D}(r,r'|E)$
of Eq.~(\ref{eq:Schro2}):
\begin{equation}
\sigma_{\textrm{tot}}=\frac{2\pi\alpha^{2}}{M_{B}s}\im{\left[\mathcal{G}^{\dagger}\mathcal{D}(0,0|E)\mathcal{G}\right]}.
\end{equation}
Here $\mathcal{G}=\left(g_{X},g_{Y}\right)^{\dagger}$, and the superscript
$\dagger$ means the hermitian conjugation. In the two-channel case
the Green's function is a matrix $2\times2$ and satisfies the equation
\begin{equation}
\left(p_{r}^{2}+M_{B}\mathcal{V}+\frac{l(l+1)}{r^{2}}-\mathcal{K}^{2}\right)\mathcal{D}(r,r'|E)=\frac{1}{rr'}\delta(r-r')\,.
\end{equation}
The Green's function can be written in terms of the regular solutions
$\Psi_{1R}$, $\Psi_{2R}$ and the non-regular solutions $\Psi_{1N}$,
$\Psi_{2N}$ of Eq.~(\ref{eq:Schro2}):
\begin{align}
\mathcal{D}(r,r'|E) & =k_{X}\left(\Psi_{1R}(r)\Psi_{1N}^{T}(r')\theta(r'-r)+\Psi_{1R}(r')\Psi_{1N}^{T}(r)\theta(r-r')\right)\nonumber \\
 & +k_{Y}\left(\Psi_{2R}(r)\Psi_{2N}^{T}(r')\theta(r'-r)+\Psi_{2R}(r')\Psi_{2N}^{T}(r)\theta(r-r')\right).
\end{align}
Corresponding upper ($u_{1N}$, $u_{2N}$) and lower ($v_{1N}$, $v_{2N}$)
components of non-regular solutions of the Schrödinger equation have
the asymptotics 
\begin{align}
 & u_{1N}(r\to\infty)=\frac{1}{kr}\chi_{X}^{+}\,,\qquad\lim_{r\to\infty}r\,v_{1N}=0\,,\nonumber \\
 & \lim_{r\to\infty}r\,u_{2N}=0\,,\qquad v_{2N}(r\to\infty)=\frac{1}{kr}\chi_{Y}^{+}\,.
\end{align}

\begin{figure}
\centering
\includegraphics[totalheight=6cm]{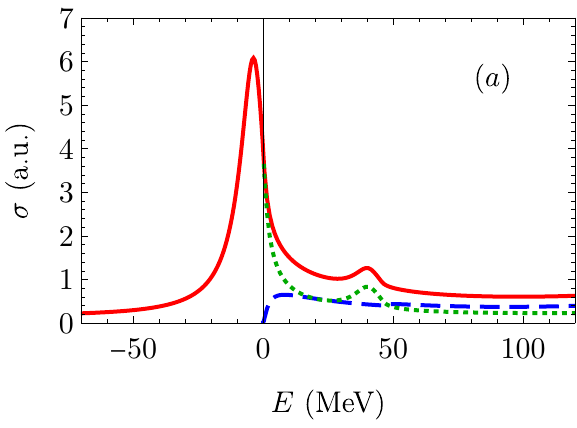}

\includegraphics[totalheight=6cm]{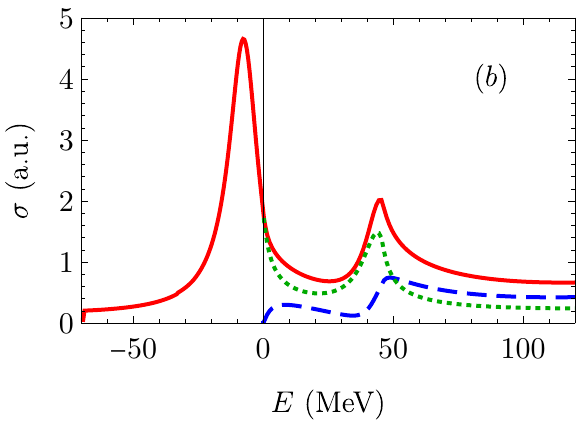}\hspace*{\fill}\includegraphics[totalheight=6cm]{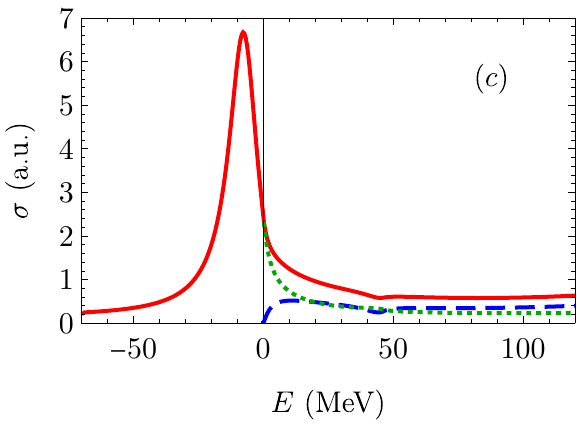}

\caption{The cross sections of hadron production for $a=\unit[1.5]{fm}$,
$U_{XX}=U_{YY}=\unit[\left(-525-10i\right)]{MeV}$, $g_{X}=\unit[1]{fm}$,
$g_{Y}=\unit[0.3]{fm}$. On plot (a) off-diagonal potential $U_{XY}=0$,
on plot (b) $U_{XY}=\unit[20]{MeV}$, and on plot (c) $U_{XY}=-\unit[20]{MeV}$.
The dashed line corresponds to the elastic cross section, the dotted
line to the inelastic cross section (annihilation into other mesons),
and the solid line to the total cross section.}\label{fig:sig2tot}
\end{figure}

When both $X$ and $Y$ channels have low-energy bound states, the
cross sections are shown in Fig.~\ref{fig:sig2tot} for small values
of $\im{V_{XX}}$ and $\im{V_{YY}}$. For $V_{XY}=0$, the bound states
in both channels manifest itself only as peaks in the inelastic and
total cross sections (see Fig.~\ref{fig:sig2tot}a). However, for
$V_{XY}=\unit[20]{MeV}$, the bound state in the $Y$ channel affects
the elastic cross sections as well. As a result, the cross section
$\sigma_{\textrm{el},X}$ decreases below the threshold of $Y$ state
production, and the peak in the total cross section at energy near
$\unit[50]{MeV}$ becomes more pronounced (see Fig.~\ref{fig:sig2tot}b).
For another sign of the off-diagonal potential, $V_{XY}=-\unit[20]{MeV}$,
the interference between channels leads to the opposite effect. Namely,
the peak in the cross section close to the threshold of $Y$ state
production disappears (see Fig.~\ref{fig:sig2tot}c).

\subsection{Three channels}

The most interesting case from the experimental point of view is the
process in which the final state consists of hadron pairs in three
channels with the same quantum numbers $J^{PC}=1^{--}$ but different
spin states. To be specific, we consider the states $X=B\bar{B}$,
$Y=\left(B^{*}\bar{B}-B\bar{B}^{*}\right)/\sqrt{2}$ and $Z=B^{*}\bar{B}^{*}$.
As mentioned above, different spin of $B$ and $B^{*}$ mesons leads
to slightly different interaction potentials in all three channels.
The transitions between channels are possible, because the quantum
numbers $J^{PC}$ of all states are identical.

The approach used for description of three-channel problem is a simple
generalization of that for two-channel case. The corresponding Schrödinger
equation reads
\begin{equation}
\left(p_{r}^{2}+M_{B}\mathcal{V}+\frac{l(l+1)}{r^{2}}-\mathcal{K}^{2}\right)\Psi(r)=0\,,\quad\mathcal{K}^{2}=\begin{pmatrix}k_{X}^{2} & 0 & 0\\
0 & k_{Y}^{2} & 0\\
0 & 0 & k_{Z}^{2}
\end{pmatrix},\quad\mathcal{V}=\begin{pmatrix}V_{XX} & V_{XY} & V_{XZ}\\
V_{XY} & V_{YY} & V_{YZ}\\
V_{XZ} & V_{YZ} & V_{ZZ}
\end{pmatrix},\label{eq:Schro3}
\end{equation}
where $k_{Z}=\sqrt{M_{B}\left(E-2\Delta\right)}$, and other momenta
are defined in the previous subsection. In Eq.~(\ref{eq:Schro3}),
the wave function $\Psi=\left(u,v,w\right)^{T}$ consists of radial
wave functions for the $X$, $Y$ and $Z$ states. As above, the angular
momentum of all states is $l=1$.

The cross sections of pair production in each state are
\begin{align}
 & \sigma_{X}=\frac{2\pi\beta_{X}\alpha^{2}}{s}\left|g_{X}u_{1R}^{(1)}(0)+g_{Y}v_{1R}^{(1)}(0)+g_{Z}w_{1R}^{(1)}(0)\right|^{2},\nonumber \\
 & \sigma_{Y}=\frac{2\pi\beta_{Y}\alpha^{2}}{s}\left|g_{X}u_{2R}^{(1)}(0)+g_{Y}v_{2R}^{(1)}(0)+g_{Z}w_{2R}^{(1)}(0)\right|^{2},\nonumber \\
 & \sigma_{Z}=\frac{2\pi\beta_{Z}\alpha^{2}}{s}\left|g_{X}u_{3R}^{(1)}(0)+g_{Y}v_{3R}^{(1)}(0)+g_{Z}w_{3R}^{(1)}(0)\right|^{2},\label{eq:sig3}
\end{align}
where $\beta_{Z}=k_{Z}/M_{B}$ and $g_{X}$, $g_{Y}$, $g_{Z}$~are
some constants related to the amplitudes of the corresponding state
production at small distances. The functions $u_{iR}$, $v_{iR}$
and $w_{iR}$ are the components of three regular solutions $\Psi_{iR}$,
having the asymptotics at $r\to\infty$
\begin{align}
 & \Psi_{1R}=\frac{1}{2ik_{X}r}\left(S_{11}\chi_{X}^{+}-\chi_{X}^{-},\,S_{12}\chi_{Y}^{+},\,S_{13}\chi_{Z}^{+}\right)^{T},\nonumber \\
 & \Psi_{2R}=\frac{1}{2ik_{Y}r}\left(S_{21}\chi_{X}^{+},\,S_{22}\chi_{Y}^{+}-\chi_{Y}^{-},\,S_{23}\chi_{Z}^{+}\right)^{T},\nonumber \\
 & \Psi_{3R}=\frac{1}{2ik_{Z}r}\left(S_{31}\chi_{X}^{+},\,S_{32}\chi_{Y}^{+},\,S_{33}\chi_{Z}^{+}-\chi_{Z}^{-}\right)^{T},\nonumber \\
 & \chi_{X}^{\pm}=\exp\left[\pm i\left(k_{X}r-\pi/2\right)\right],\quad\chi_{Y}^{\pm}=\exp\left[\pm i\left(k_{Y}r-\pi/2\right)\right],\nonumber \\
 & \chi_{Z}^{\pm}=\exp\left[\pm i\left(k_{Z}r-\pi/2\right)\right],
\end{align}
where $S_{ij}$~are some coefficients.

In order to demonstrate how the transitions between channels affect
the energy dependence of the cross sections, let us consider a few
simple examples. We choose the parametrization of the potential in
the form of rectangular wells with the same radius but different depths:
\begin{equation}
\mathcal{V}=\begin{pmatrix}U_{XX} & U_{XY} & U_{XZ}\\
U_{XY} & U_{YY} & U_{YZ}\\
U_{XZ} & U_{YZ} & U_{ZZ}
\end{pmatrix}\theta(a-r)\,,
\end{equation}
where the matrix elements $U_{ij}$ and the radius $a$~are some
constants. Below we present the numerical results obtained within
this model for $a=\unit[1.5]{fm}$, $g_{X}=\unit[1]{fm}$, $g_{Y}=\unit[0.3]{fm}$,
$g_{Z}=\unit[0.3]{fm}$, and various depths of the potential wells.

\begin{figure}
\centering
\includegraphics[totalheight=6cm]{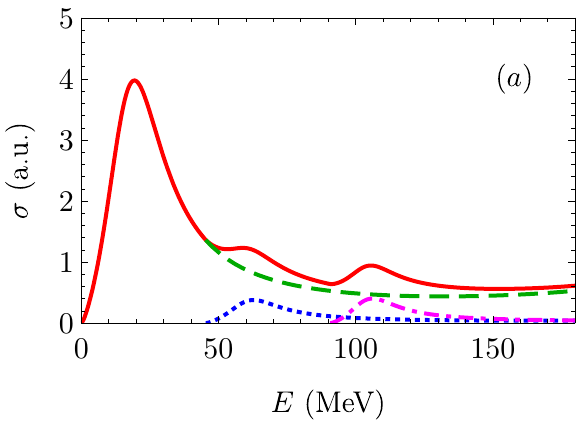}\hspace*{\fill}\includegraphics[totalheight=6cm]{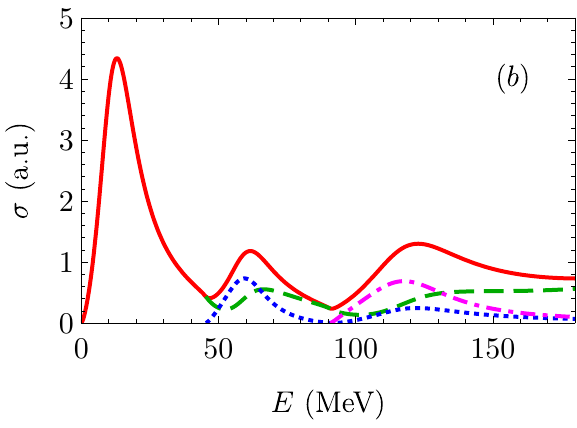}

\caption{The energy dependence of the cross sections $\sigma_{X}$ (dashed
line), $\sigma_{Y}$ (dotted line), $\sigma_{Z}$ (dash-dotted line)
and their sum (solid line). The parameters of the model are $U_{XX}=U_{YY}=U_{ZZ}=-\unit[490]{MeV}$,
$a=\unit[1.5]{fm}$, $g_{X}=\unit[1]{fm}$, $g_{Y}=\unit[0.3]{fm}$,
$g_{Z}=\unit[0.3]{fm}$. On plot (a) off-diagonal potentials $U_{XY}=U_{XZ}=U_{YZ}=0$,
on plot (b) $U_{XY}=U_{XZ}=U_{YZ}=\unit[20]{MeV}$.}\label{fig:sig3-1}
\end{figure}

First consider the case when there are low-energy virtual states in
all three channels. If~the transitions between channels are absent,
then the peaks corresponding to these virtual states are seen, but
the contributions of the $Y$ and $Z$ states to the total cross section
are small (see Fig.~\ref{fig:sig3-1}a). However, for the off-diagonal
potentials $U_{XY}=U_{XZ}=U_{YZ}=\unit[20]{MeV}$, the peaks and dips
in the cross sections are more pronounced (see Fig.~\ref{fig:sig3-1}b).
Because of the transitions between channels, the cross section $\sigma_{X}$
has strong energy dependence in the vicinity of $Y$ and $Z$ thresholds.
Moreover, the cross section $\sigma_{Y}$ becomes almost zero at the
threshold of $Z$ state. For the opposite signs of the mixing potentials
the dips in the cross sections disappear and the peaks become less
pronounced. If there are virtual states in each channel, then the
energy dependence of the cross sections in the vicinity of each threshold
looks similar to that for the two-channels problem, see Fig.~\ref{fig:sig2-1}.

\begin{figure}
\centering
\includegraphics[totalheight=6cm]{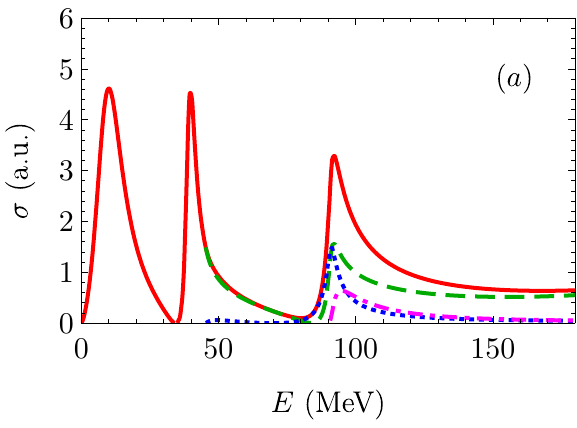}\hspace*{\fill}\includegraphics[totalheight=6cm]{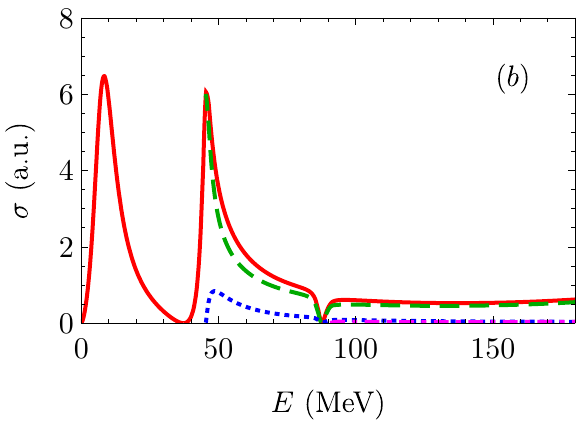}

\caption{The energy dependence of the cross sections $\sigma_{X}$ (dashed
line), $\sigma_{Y}$ (dotted line), $\sigma_{Z}$ (dash-dotted line)
and their sum (solid line). The parameters of the model are $U_{XX}=-\unit[490]{MeV}$,
$U_{YY}=U_{ZZ}=-\unit[525]{MeV}$, $a=\unit[1.5]{fm}$, $g_{X}=\unit[1]{fm}$,
$g_{Y}=\unit[0.3]{fm}$, $g_{Z}=\unit[0.3]{fm}$. On plot (a) off-diagonal
potentials $U_{XY}=U_{XZ}=U_{YZ}=\unit[20]{MeV}$, on plot (b) $U_{XY}=\unit[20]{MeV}$,
$U_{XZ}=\unit[-20]{MeV}$, $U_{YZ}=0$.}\label{fig:sig3-2}
\end{figure}

Then consider a virtual state in the $X$ channel and low-energy bound
states in the other channels. If there are no transitions between
channels, then these bound states are not observable. However, for
$U_{XY}=U_{XZ}=U_{YZ}=\unit[20]{MeV}$, these states can decay into
lighter particles. Namely, the bound state in the $Y$ channel can
pass into the $X$ channel, and the bound state in the $Z$ channel
can pass into both $X$ and $Y$ channels. As a result, sharp peaks
occur close to the thresholds of $Y$ and $Z$ states, and the total
cross section falls almost to zero below these peaks (see Fig.~\ref{fig:sig3-2}a).
Note that the energy dependence of the cross sections may change drastically
for other values of mixing potentials. For instance, the cross section
energy dependence for $U_{XY}=\unit[20]{MeV}$, $U_{XZ}=\unit[-20]{MeV}$,
$U_{YZ}=0$ is shown in Fig.~\ref{fig:sig3-2}b. The direct transition
between the $Y$ and $Z$ channels is now forbidden, hence there is
no pronounced peak corresponding to the bound state in the $Z$ channel.
Nevertheless, the total cross section has a sharp dip close to the
threshold of the $Z$ state production.

Obviously, other combinations of real and virtual states in the potentials,
as well as the signs of the mixing potentials and the constants $g_{X}$,
$g_{Y}$, $g_{Z}$, are also possible. For each set of parameters,
the cross sections have their own peculiarities. We focused our attention
on such parameters which lead to the energy dependence of the cross
sections similar to the experimental one. The detailed description
of the experimental data within our model is performed in the next
section.

\section{Production of $B^{(*)}\bar{B}^{(*)}$ near the thresholds}\label{sec:Production-of-}

\begin{figure}
\centering
\includegraphics[totalheight=5.7cm]{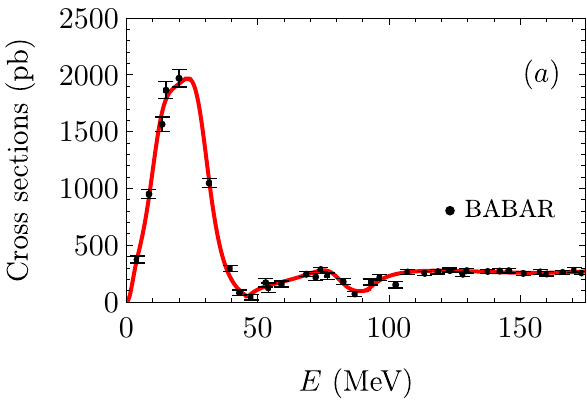}\hspace*{\fill}\includegraphics[totalheight=5.7cm]{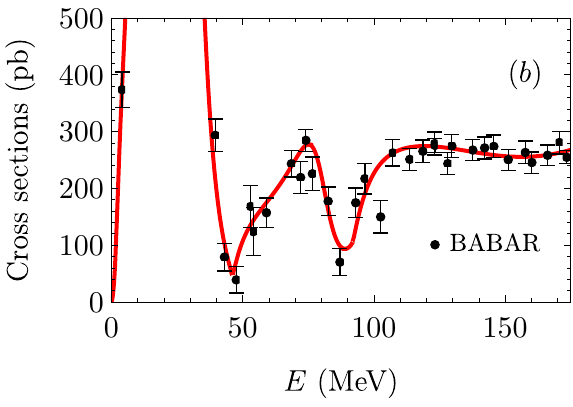}

\includegraphics[totalheight=5.7cm]{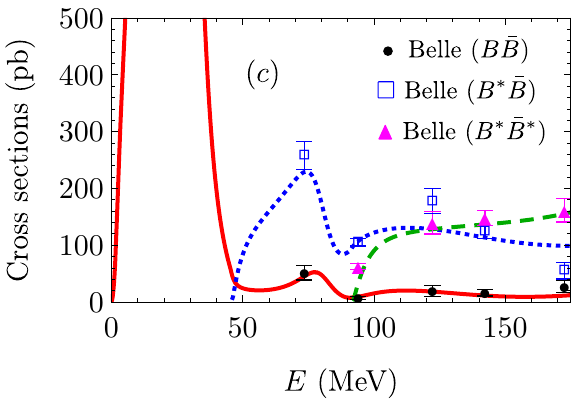}

\caption{The energy dependence of the cross sections of $B$ mesons production.
The total cross section $\sigma_{X}+\sigma_{Y}+\sigma_{Z}$ is shown
on plots (a) and (b) with the solid line. On plot (c) the exclusive
cross sections $\sigma_{X}$ (solid line), $\sigma_{Y}$ (dotted line),
and $\sigma_{Z}$ (dashed line) are shown. Experimental data for the
total cross section were recalculated in Ref.~\citep{Dong2020} using
the data~\citep{Aubert2009}. Experimental data for the exclusive
cross sections are taken from Refs.~\citep{Mizuk2021,Bertemes2023}.}\label{fig:sigBB}
\end{figure}

Precise measurement of $B\bar{B}$, $B\bar{B}^{*}$, $B^{*}\bar{B}$,
$B^{*}\bar{B}^{*}$ pair production cross sections in $e^{+}e^{-}$
annihilation near the thresholds of the corresponding processes~\citep{Mizuk2021,Bertemes2023},
as well as the total cross section of these processes~\citep{Aubert2009}
demonstrate a non-trivial energy dependence. There are peaks with
very unusual shape and deep dips in these cross sections (see Fig~\ref{fig:sigBB}).
In this section, we show that our approach described above completely
explains such a behavior.

As mentioned above, the states $X=B\bar{B}$, $Y=\left(B^{*}\bar{B}-B\bar{B}^{*}\right)/\sqrt{2}$
and $Z=B^{*}\bar{B}^{*}$ have the same quantum numbers $J^{PC}=1^{--}$
and isospin $I=0$. These quantum numbers correspond to the angular
momentum $l=1$ of meson pairs. The spin of $B^{*}\bar{B}^{*}$ pair
can be $S=0$ or $S=2$. Since in the present paper we do not discuss
the angular distributions of produced $B^{*}$ and $\bar{B}^{*}$
mesons, below we talk about the sum of cross sections with spins $S=0$
and $S=2$. The thresholds of $X$, $Y$, and $Z$ channels are close
to each other, so the transition amplitudes between these channels
are essential.

The isotopic invariance violating effects, such as the Coulomb interaction
between charged mesons and the mass difference of charged and neutral
mesons, result in slight shift of peak positions in the cross sections
of charged and neutral meson production. This effect has been studied
in detail in Refs.~\citep{Milstein2021,Bondar2022}. Since the main
goal of the present paper is to study the influence of transitions
between several channels on the energy dependence of the cross sections,
we do not discuss the effects of isotopic invariance violation. However,
we have checked that these effects only slightly modify the energy
dependence of the cross sections.

The exclusive cross sections of various final states ($X$, $Y$,
and $Z$) production can be calculated directly by means of Eq.~(\ref{eq:sig3}),
where the wave functions are the corresponding solutions of the Schrödinger
equation~(\ref{eq:Schro3}) for the three-channel case. For the description
of experimental data, we use the parametrization of diagonal and off-diagonal
potentials as the rectangular wells
\begin{equation}
V_{ij}(r)=U_{ij}\cdot\theta(a_{ij}-r)\,,\qquad i,j=X,Y,Z\,.\label{eq:pot}
\end{equation}
Of course, the model potentials~(\ref{eq:pot}) do not match the
real interaction potentials. However, as explained above, if there
are low-energy real or virtual states, the specific form of the potentials
are not very important to reproduce the experimental data near the
thresholds. The use of model potentials in the form~(\ref{eq:pot})
simplifies significantly calculation of the cross sections in the
multichannel case.

Our analysis shows that it is possible to neglect the imaginary parts
of $U_{ij}$ since the probabilities for $B^{(*)}\bar{B}^{(*)}$ pairs
to annihilate into lighter particles are small. The quantities $U_{ij}$,
$a_{ij}$, as well as the complex constants $g_{X}$, $g_{Y}$, and
$g_{Z}$, are the fitting parameters of our model. The values of these
parameters, providing the best agreement between our predictions and
the experimental data, are listed in Table~\ref{tab:params}. The
energy dependence of the total cross section $\sigma_{\textrm{tot}}=\sigma_{X}+\sigma_{Y}+\sigma_{Z}$,
as well as the exclusive cross sections $\sigma_{X}$, $\sigma_{Y}$,
and $\sigma_{Z}$, is shown in Fig.~\ref{fig:sigBB}. Note that there
are other sets of potential parameters that provide good agreement
with experimental data. However, we rejected them for several reasons.
Firstly, the values of the diagonal potentials differed from each
other by several times. Secondly, the potential radii were either
very small or very large compared to 1~fm. Thirdly, the off-diagonal
potentials were very large. Under these restrictions, the potential
parameters specified in the Table~\ref{tab:params} are determined
with an accuracy of a few percents.

\setlength{\tabcolsep}{0.4em}

\begin{table}
\centering
\begin{tabular}{|l|>{\centering}m{2cm}|>{\centering}m{2cm}|>{\centering}m{2cm}|>{\centering}m{2cm}|>{\centering}m{2cm}|>{\centering}m{2cm}|}
\hline 
 & $V_{XX}$ & $V_{YY}$ & $V_{ZZ}$ & $V_{XY}$ & $V_{XZ}$ & $V_{YZ}$\tabularnewline
\hline 
$U_{ij}\,(\mathrm{MeV})$ & $-613.1$ & $-360.6$ & $-586.7$ & $26.7$ & $20$ & $78.6$\tabularnewline
$a_{ij}\,(\mathrm{fm})$ & $1.361$ & $1.804$ & $1.809$ & $0.953$ & $2.819$ & $2.209$\tabularnewline
\hline 
$g_{i}\,(\mathrm{fm})$ & \multicolumn{2}{c|}{$g_{X}=0.118$} & \multicolumn{2}{c|}{$g_{Y}=-0.004+0.217\,i$} & \multicolumn{2}{c|}{$g_{Z}=-0.6+0.193\,i$}\tabularnewline
\hline 
\end{tabular}

\caption{Parameters of the model describing $B$ mesons production in $e^{+}e^{-}$
annihilation.}\label{tab:params}
\end{table}

A peak in the total cross section corresponding to $\Upsilon(4S)$
resonance is observed at an energy around $\unit[20]{MeV}$. This
peak is the result of the production of $B\bar{B}$ pair near the
threshold in the presence of a virtual state in this channel. There
is also a peak at an energy around $\unit[75]{MeV}$ corresponding
to $B\bar{B}^{*}$ and $B^{*}\bar{B}$ pair production, as well as
a broad peak above $\unit[100]{MeV}$ associated with the production
of $B^{*}\bar{B}^{*}$ pairs. It turned out, however, that the positions
and shapes of the latter two peaks depend strongly on the magnitude
of the off-diagonal potentials responsible for the transitions between
the channels. Although the magnitudes of these potentials are much
smaller than the diagonal potentials (see Table~\ref{tab:params}),
these magnitudes are comparable to the energies of real or virtual
states, which are also small compared to the diagonal potentials.
This is why the transitions between channels significantly affect
the energy dependence of the cross sections. The transitions between
channels are also responsible for the sharp dips in the cross section
at energies near $\unit[45]{MeV}$ and $\unit[90]{MeV}$. Neglecting
the off-diagonal potentials, we obtain that the dips in the cross
section practically disappear. It is seen that our model describes
well the main features of the total cross sections.

Our model explains also the unusual energy dependence of the exclusive
cross sections. For instance, there is a peak in $\sigma_{Y}$ below
the threshold of $B^{*}\bar{B}^{*}$ production (see Fig.~\ref{fig:sigBB}c).
This peak is a manifestation of $B^{*}\bar{B}^{*}$ bound state, because
this state can decay into the $Y$ channel. Thus, such a decay is
possible solely due to the transitions between $Y$ and $Z$ channels.
It is a very non-trivial effect, and its detailed study seems to be
very important. Apparently, similar phenomena are also observed in
the cross sections for the production of various $D$ mesons, but
this problem requires special investigation.

For zero off-diagonal potentials, the energy of $B^{*}\bar{B}^{*}$
bound state is $\unit[67]{MeV}$ that is $\unit[25]{MeV}$ below the
$B^{*}\bar{B}^{*}$ threshold. We expect that this bound state can
also manifest itself in the inelastic processes, where non-$B$-meson
final states are produced. The width of the peak in the corresponding
cross sections is expected to be about $\unit[20]{MeV}$.

\section{Conclusion}

In our work, the production of hadron pairs near the thresholds is
discussed, when there are low-energy bound or virtual states of produced
particles. The consideration is based on the effective potential approach,
which takes into account the interaction between hadrons in the final
state. Particular attention is paid to the case when there are several
reaction channels with nonzero transition amplitudes between them.
It is shown that the energy dependence of the cross sections is very
sensitive to the off-diagonal potentials, though the latter are small
as compared to the diagonal ones. In particular, a narrow resonance
below the threshold in one channel can lead to broad peaks in other
channels.

Using our approach we have explained all available experimental data
for the cross sections of $B\bar{B}$, $B^{*}\bar{B}$, $B\bar{B}^{*}$
and $B^{*}\bar{B}^{*}$ pair production in $e^{+}e^{-}$ annihilation.
It is shown that the non-trivial shape of the peaks in the cross sections,
as well as the sharp dips, are the result of the transitions between
different channels.

\end{document}